\documentclass[12pt]{iopart}

\usepackage{harvard}
\usepackage{graphicx}   % alternative graphics specificati.jpgons
\usepackage{hyperref}   %link

\hypersetup{
    bookmarks=true,         % show bookmarks bar?
    unicode=false,          % non-Latin characters in Acrobat?s bookmarks
    colorlinks=true,       % false: boxed links; true: colored links
    linkcolor=blue,          % color of internal links (change box color with linkbordercolor)
    citecolor=blue,        % color of links to bibliography
    filecolor=blue,      % color of file links
    urlcolor=blue           % color of external links
}

\begin{document}

\title[Nucleosynthesis in early rotating massive stars]{Nucleosynthesis in early rotating massive stars and chemical composition of CEMP stars}

\author{A Choplin$^{1}$ and R Hirschi$^{2,3,4}$ 
}

\address{$^1$ Department of Physics, Faculty of Science and Engineering, Konan University, 8-9-1 Okamoto, Kobe, Hyogo 658-8501, Japan}

\address{$^2$ Astrophysics Group, Lennard-Jones Labs 2.09, Keele University, ST5 5BG, Staffordshire, UK}

\address{$^3$ Kavli Institute for the Physics and Mathematics of the Universe (WPI), University of Tokyo, 5-1-5 Kashiwanoha, Kashiwa, 277-8583, Japan}

\address{$^4$ UK Network for Bridging the Disciplines of Galactic Chemical Evolution (BRIDGCE)}

\ead{arthur.choplin@konan-u.ac.jp}

\begin{abstract}
The first massive stars triggered the onset of chemical evolution by releasing the first metals (elements heavier than helium) in the Universe. The nature of these stars and how the early chemical enrichment took place is still largely unknown. 
Rotational-induced mixing in the stellar interior can impact the nucleosynthesis during the stellar life of massive stars and lead to stellar ejecta having various chemical compositions.
We present low and zero-metallicity 20, 25 and 40 $M_{\odot}$ stellar models with various initial rotation rates and assumptions for the nuclear reactions rates. With increasing initial rotation, the yields of light (from $\sim$ C to Al) and trans-iron elements are boosted. The trans-iron elements (especially elements heavier than Ba) are significantly affected by the nuclear reaction uncertainties. The chemical composition of the observed CEMP (carbon-enhanced metal-poor) stars CS29528-028 and HE0336+0113 are consistent with the chemical composition of the material ejected by a fast rotating 40~$M_{\odot}$ model.
\end{abstract}

\vspace{2pc}

\section{Introduction}

The nature of the short-lived first generations of massive stars, that released the first metals (elements heavier than helium) in the Universe, is still largely unknown \cite{nomoto13,karlsson13}. 
The nature of these stars can be probed indirectly by investigating the origin of low-mass metal-poor stars observed in our neighborhood, some of which are almost as old as the Universe \cite{beers05,frebel15}. The material forming these living fossils is thought to have been inherited from the ejecta of one or very few previous massive stars. 
The great variety of chemical composition patterns of observed low-mass metal-poor stars suggests that the characteristics and therefore the yields (chemical composition of the ejected material) of the first massive stars were very diverse \cite{umeda03,tominaga14,maeder15a,yoon16,placco16b,choplin16}. 
Among the most important parameters impacting the yields of massive stars are the initial mass, rotation, magnetic fields or binarity \cite{hirschi07,heger10,yoon05,yoon12,maeder12}. Rotation in the first massive stars was found to be an interesting ingredient so as to account for the peculiar abundance patterns of many metal-poor stars \cite{meynet06,takahashi14,banerjee19}. 
Low and zero metallicity massive stellar models including rotation and full nucleosynthesis were calculated recently \cite{frischknecht16,limongi18,choplin18,banerjee19} but there is still a clear need of computing more of these models with different initial assumptions to obtain a clearer picture of the nucleosynthesis in the first massive stars.

Here, we investigate the impact of rotation on the nucleosynthesis of light and heavy elements in low and zero metallicity $20-40$~$M_{\odot}$ stars including rotation. 
We then discuss the possibility of the existence of fast rotating massive stars in the early Universe by confronting our model yields to the chemical composition of observed metal-poor stars.

\section{Low and zero metallicity massive stellar models including rotation}\label{models}

\begin{table}
\caption{Characteristics of the models computed. From left to right: initial mass, metallicity and corresponding [Fe/H], $v_{\rm ini}/v_{\rm crit}$, initial equatorial velocity and the set of nuclear reaction rates used (see text for details). \label{table:1}}
\begin{indented}
\item[]\begin{tabular}{@{}ccccccc}
\hline % inserts double horizontal lines
\hline % inserts single horizontal line
 $M_{\rm ini}$ &  Z  & [Fe/H] & $v_{\rm ini}/v_{\rm crit}$ & $v_{\rm ini}$ & Nuclear rates & Model label \\ % table heading
$[M_{\odot}$] & &   & & [km~s$^{-1}$] & \\
\hline
20			& 0 		   & $-\infty$ &0.4 &479& set A & 20z0v4A	\\
\hline % inserts single horizontal line
20			& $10^{-5}$ & $-3.8$ &0    & 0 & set A & 20z5v0A	\\
20			& $10^{-5}$ & $-3.8$ &0.1 & 88& set A & 20z5v1A	\\
20			& $10^{-5}$ & $-3.8$ &0.2 &188& set A & 20z5v2A	\\
20			& $10^{-5}$ & $-3.8$ &0.3 &276& set A & 20z5v3A	\\
20			& $10^{-5}$ & $-3.8$ &0.4 &364& set A & 20z5v4A	\\
20			& $10^{-5}$ & $-3.8$ &0.5 &454& set A & 20z5v5A	\\
20			& $10^{-5}$ & $-3.8$ &0.6 &547& set A & 20z5v6A	\\
20			& $10^{-5}$ & $-3.8$ &0.7 &644& set A & 20z5v7A	\\
\hline
20			& $10^{-3}$ & $-1.8$ &0.4 &  306	& set A & 20z3v4A\\
25			& $10^{-3}$ & $-1.8$ &0    & 0	& set A & 25z3v0A	\\
25			& $10^{-3}$ & $-1.8$ &0.4 &320& set A & 25z3v4A	\\
25			& $10^{-3}$ & $-1.8$ &0.7 &560& set A & 25z3v7A 	\\
25			& $10^{-3}$ & $-1.8$ &0.7 &560& set B & 25z3v7B	\\
25			& $10^{-3}$ & $-1.8$ &0.7 &560& set C & 25z3v7C	\\
40			& $10^{-3}$ & $-1.8$ &0.8 &704& set C & 40z3v8C	\\
\hline
\end{tabular}
\end{indented}
\end{table}

   \begin{figure}
   \centering
   \begin{minipage}[c]{.2\linewidth}
       \includegraphics[scale=0.41]{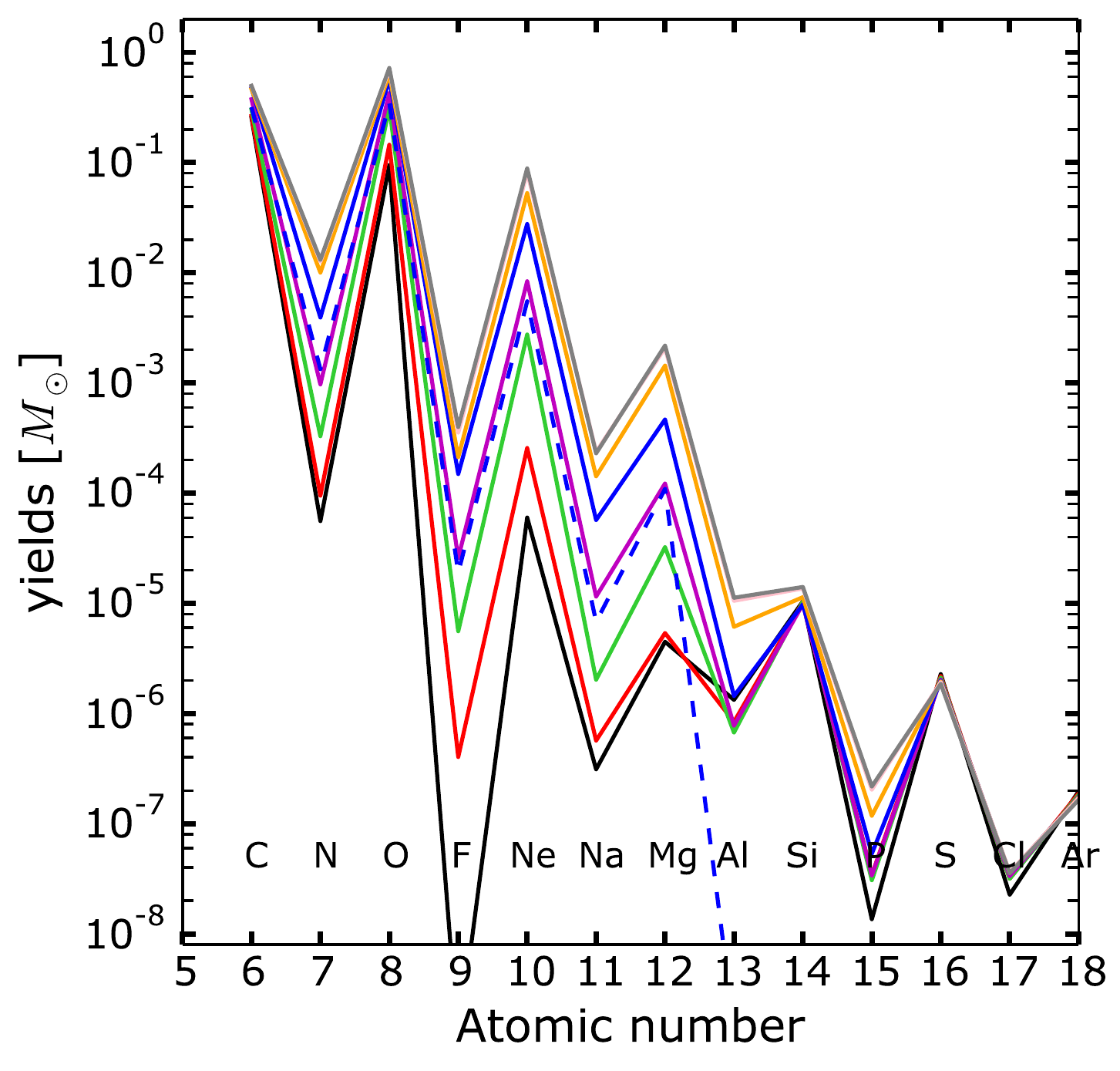}
   \end{minipage}
   \hspace{2.7cm}
   \begin{minipage}[c]{.6\linewidth}
       \includegraphics[scale=0.41]{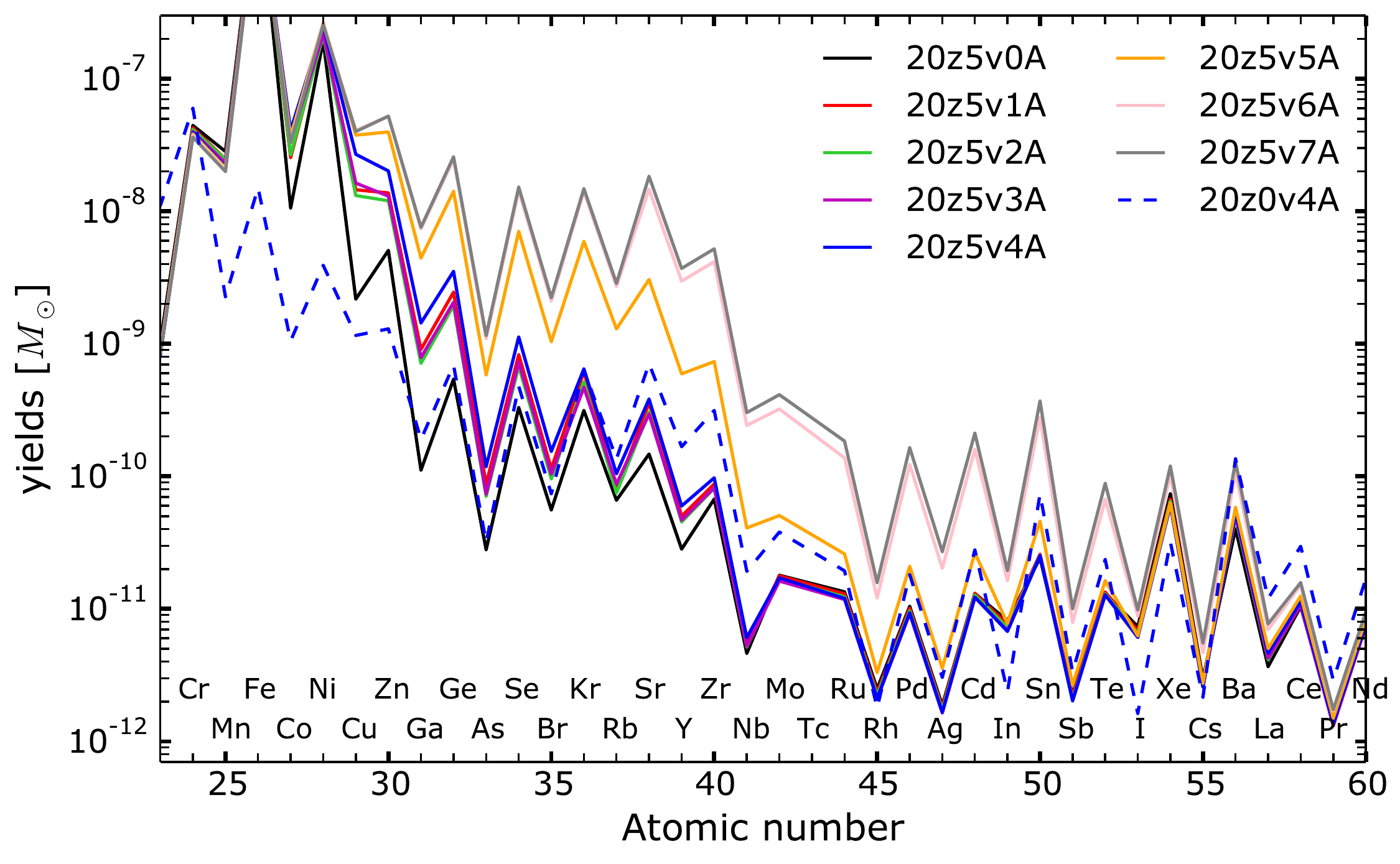}
   \end{minipage}
   \caption{
   Yields of light (\textit{left panel}) and heavy (\textit{right panel}) elements for the 20 $M_{\odot}$ models at metallicity $Z=10^{-5}$. The 20 $M_{\odot}$, $Z=0$ model is also shown (dashed line). Its yields for heavy elements (\textit{right panel}) are scaled up by a factor of $10^6$ .
   }
\label{fig1}
    \end{figure}

It has been shown in different works that during the core helium-burning stage of rotating massive stars, the exchanges of material between the H- and He-burning zones and triggered by rotation-induced mixing can boost the production of light elements such as nitrogen and heavy elements such as strontium or barium \cite{meynet02b,pignatari08,ekstrom08,frischknecht12,takahashi14}. 
 
Here we computed the models shown in Table~\ref{table:1} until the onset of core oxygen-burning with the GENEC stellar evolution code \cite{eggenberger08}. The nucleosynthesis is followed consistently during the evolution with a nuclear network of 737 isotopes. These models are computed with the same physical ingredients as in Choplin et al. (2018)\footnote{In the present paper, all models are new except 20z3v4A, 25z3v0A, 25z3v4A, 25z3v7A and 25z3v7B, cf. \cite{choplin18}.}. The set A of nuclear reactions is the default one \cite{choplin18}, in particular with $^{17}$O($\alpha,\gamma$) from Best et al. (2013). The set B is similar but with the $^{17}$O($\alpha,\gamma$) reaction rate divided by 10. 
This rate still suffer large uncertainties at temperatures relevant for the s-process in massive stars. At $T = 0.2 - 0.3$ GK (approximate temperature in the helium burning cores of massive stars), the rate could be $2-3$ dex smaller than the recommended rate \cite{taggart19}. 
For the set C, we took the nuclear reactions rates that favour the most the production of s-elements: $^{22}$Ne($\alpha, \gamma$) is from Iliadis et al. (2010), $^{17}$O($\alpha,\gamma$) is from Taggart et al. (2019), $^{17}$O($\alpha,n$) is from Caughlan \& Fowler (1988) and $^{22}$Ne($\alpha$,$n$) is from Angulo et al. (1999).

Figure~\ref{fig1} shows the integrated yields of models at metallicity $Z=10^{-5}$, with different $\upsilon_{\rm ini}/\upsilon_{\rm crit}$ ratios\footnote{The initial equatorial velocity is $\upsilon_{\rm ini}$ and $\upsilon_{\rm crit}$ is the initial equatorial velocity at which the gravitational acceleration is balanced by the centrifugal force. It is defined as $\upsilon_{\rm crit} = \sqrt{ \frac{2GM}{3R_{\rm pb}}}$ , where $R_{\rm pb}$ is the the polar radius at the break-up  velocity \cite{maeder00a}.} on the zero-age main-sequence. The mass cuts\footnote{At the time of the explosion, the mass cut is the mass coordinate delimiting the part of the star that is expelled from the part that is locked into the remnant.} of the models are set at the bottom of the He-shell (corresponding to a mass coordinate of $\sim 4$~$M_{\odot}$). The yields of $\sim$~C$-$Al and $\sim$~Cu$-$Te elements increase with initial rotation, as a result of the more efficient rotation-induced mixing operating between the H-burning shell and He-burning core. The Population III model (20z0v4A, dashed blue line) shows similar yields to the 20z5v4A model for C$-$Mg elements but much lower yields for heavier elements, in particular the s-elements. This is because the production of s-elements strongly depends on the initial amount of heavy seeds nuclei like iron (e.g. Prantzos et al. 1990) which is null in the 20z0v4A model.

\begin{figure}
   \centering
       \includegraphics[scale=0.53]{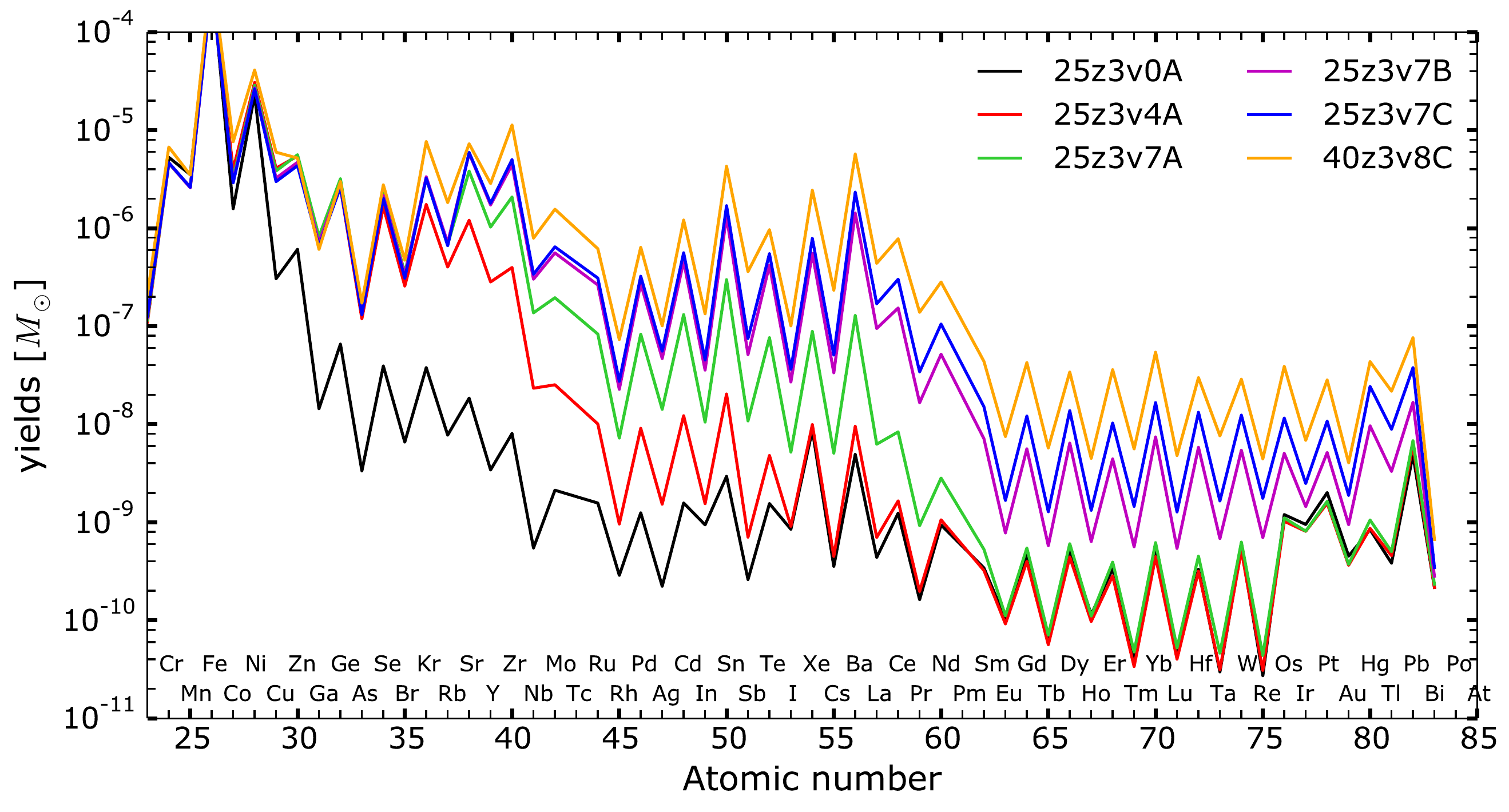}
   \caption{
   Yields of heavy elements for the 25 and 40 $M_{\odot}$ models at $Z=10^{-3}$. 
      }
\label{fig2}
    \end{figure}

At a metallicity of $Z=10^{-3}$, more seed nuclei are available and consequently more s-elements are produced (Figure~\ref{fig2}). The effect of rotation at $Z=10^{-3}$ is slightly stronger than at $Z=10^{-5}$ and elements up to $\sim$~Ba are overproduced (compare the black, red and green lines).

The nuclear reactions rates are also playing a crucial role in determining the final yields. The green, magenta and blue lines in Figure~\ref{fig2} show fast rotating 25 $M_{\odot}$, $Z=0.001$ models computed with different sets of nuclear reactions (cf. paragraph~2 of this section). The effect of nuclear uncertainties can affect the yields by about 1 dex for some elements around Nd. 

The 40z3v8C model (orange line) might present the parameters that will boost the most the production of s-elements: (1) fast rotation, (2) a mass of 40~$M_{\odot}$, that was suggested to be the mass where the production of s-elements is the highest in massive rotating stars with $Z=10^{-3}$ (Figure~7 in Choplin et al. 2018), and (3) a set of nuclear reactions favoring the s-process. Although showing the highest yields for trans-iron elements, this model does not show dramatic differences compared to the 25z3v7B (magenta line) and 25z3v7C (blue line) models.

   \begin{figure}
   \centering
       \includegraphics[scale=0.6]{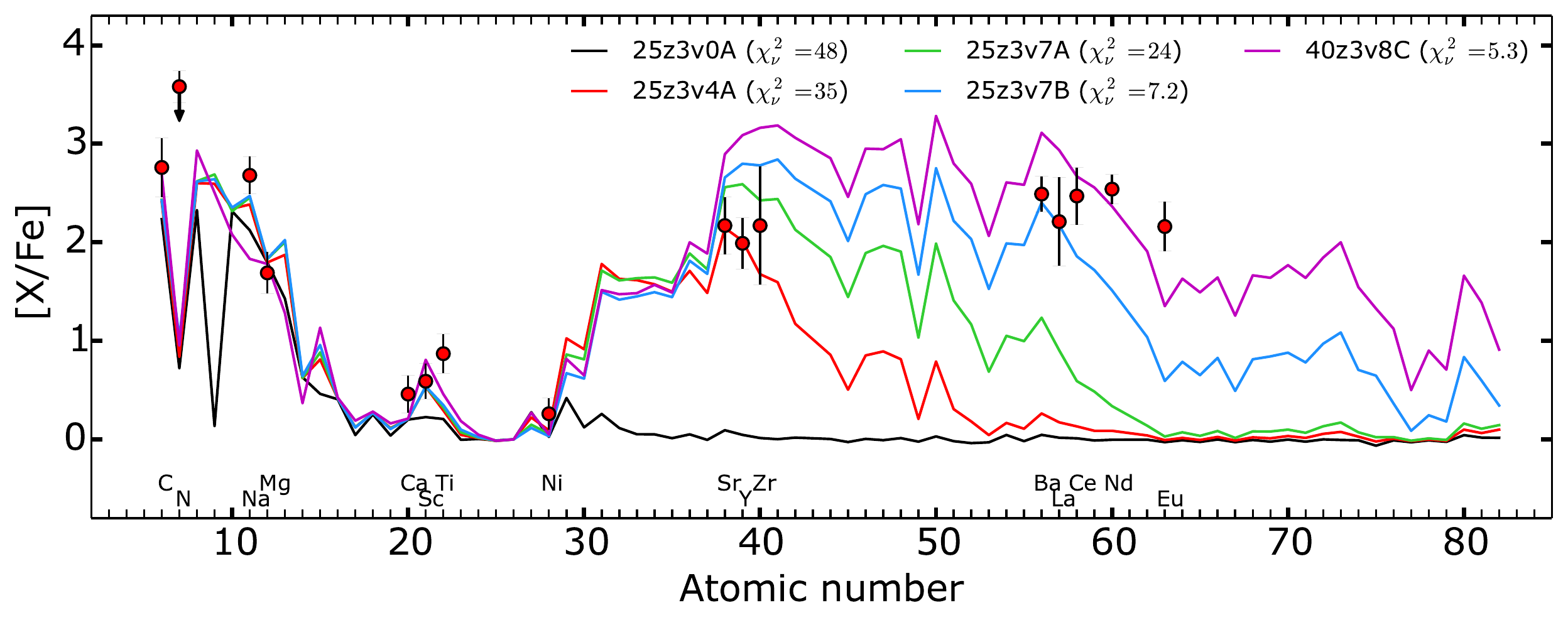}
       \includegraphics[scale=0.6]{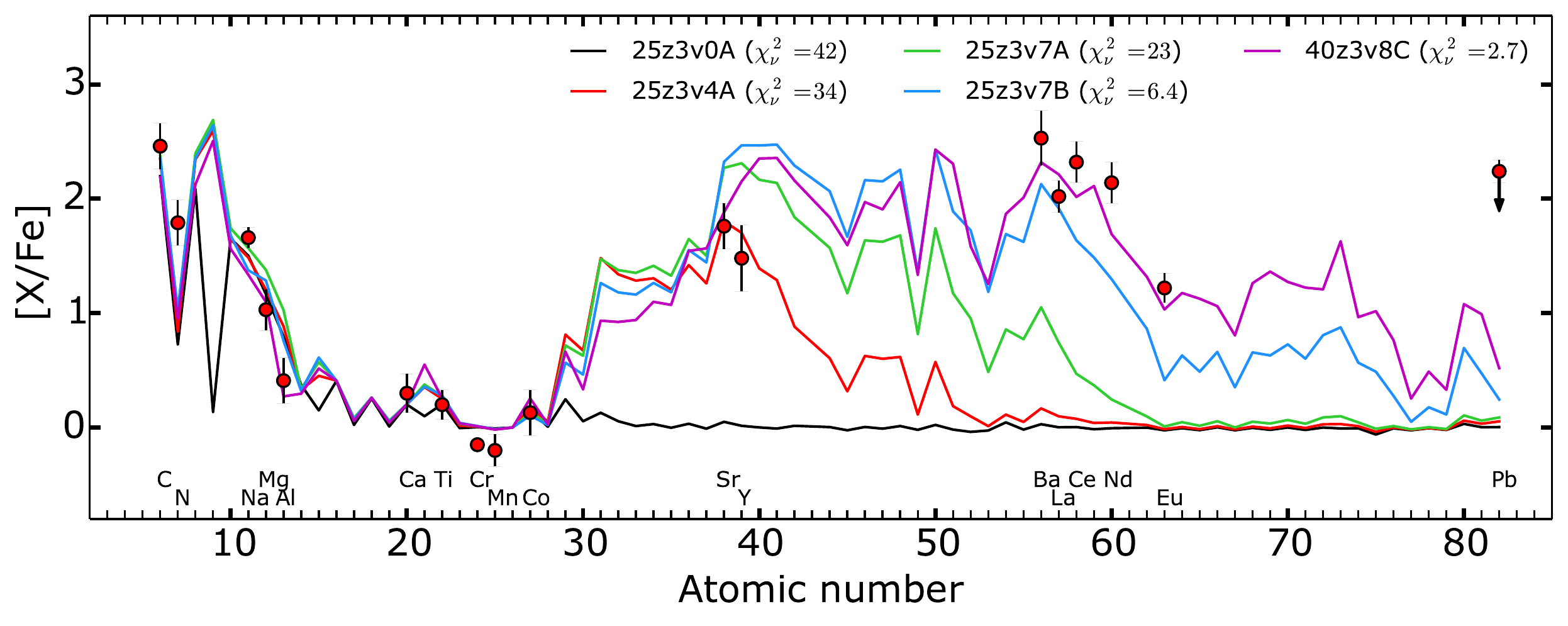}
   \caption{
   Best fits to the abundances of CS29528-028 (\textit{top panel}) and HE0336+0113 (\textit{bottom panel}). The measured stellar abundances are shown by red circles. Upper limits are indicated by arrows. The lines show the best model found for the 25 and 40 $M_{\odot}$ at $Z=10^{-3}$. The 25z3v7C model is not considered since it behaves almost as the 25z3v7B model (cf. Figure~\ref{fig2}). The reduced $\chi^2$ value is indicated in parenthesis.
   }
\label{fig3}
    \end{figure}

\section{CS29528$-$028 and HE0336+0113: descendants of rotating massive stars?}\label{mpstars}

The surface chemical composition of metal-poor stars can be compared to the chemical composition of the ejecta of low or zero-metallicity massive stars. By determining the massive model able to best match the abundances of metal-poor stars, one can infer the nature of the first massive stars. 
Here we discuss two metal-poor stars with metallicity [Fe/H] $=-2.12$ and $-2.73$ that show chemical compositions very similar to the predictions of the fast rotating models with $Z=10^{-3}$ presented in Section~\ref{models}.
For comparisons at lower metallicity, we refer to our recent work that carried out a detailed comparisons between 272 metal-poor stars with $-4<$~[Fe/H]~$<-3$ and massive stellar models including rotation computed at $Z=10^{-5}$ \cite{choplin19}.  

To fit the abundances of metal-poor stars, we vary freely the mass cut of the considered massive stellar model and also vary freely the mass of added interstellar material from $10^2$ to $10^6$~$M_{\odot}$ to the massive star ejecta. For each massive star model, we select the parameters that minimize the reduced $\chi^2$ value defined as $\chi_{\nu}^2 = \chi^2 / (N-m)$, where $N$ is the number of measured abundances for the considered metal-poor star and $m=2$ the number of free parameters to make the fit. 

CS29528-028 (Figure~\ref{fig3}, top panel) has [Fe/H] $=-2.12$ and is enriched in C, Na, Mg and trans-Fe elements \cite{aoki07,allen12}. While its abundances from C to Ni may be reproduced by any models, Sr, Y and Zr can only be reproduced if including rotation and Ba$-$Eu abundances are reasonably reproduced only by the fast rotating models 25z3v7B and 40z3v8C. When considering the 25z3v7B model, the best fit found cannot reach the high Nd and Eu values (blue line). The 40z3v8C model (magenta line) gives the lowest $\chi_{\nu}^2$ value, with a mass cut of 4.55~$M_{\odot}$.

The abundance data of HE0336+0113 (Figure~\ref{fig3}, bottom panel) is from from Cohen et al. (2006, 2013). This star has [Fe/H] $=-2.73$ with similar abundance trends than CS29528-028. Like for CS29528-028, the 40z3v8C model gives the lowest $\chi_{\nu}^2$ value but this time with a mass cut of 14.6~$M_{\odot}$, which corresponds to an explosion with a stronger amount of material falling back into the central remnant.

\section{Discussions and conclusions}

We calculated 20, 25 and 40 $M_{\odot}$ stellar models at $Z=10^{-3}$, $10^{-5}$ and 0, with various initial rotation rates and assumptions for the nuclear reactions rates. The ejecta composition of light (from $\sim$ C to Al) and trans-iron elements is affected by rotation. The nuclear reaction uncertainties significantly affect the production of trans-iron elements, especially for elements heavier than Ba.

CS29528-028 and HE0336+0113 show patterns somewhat consistent with $40$~$M_{\odot}$ fast rotating models (cf. Section~\ref{mpstars}) and therefore might suggest the existence of such massive stars in the early Universe. 
Nevertheless, some discrepancies remain, particularly for CS29528-028: accounting for the rather high [Eu/Fe] ratio requires to overproduce some lighter elements like Sr, Y and Zr. It may be that part or all of the heaviest elements (from Ba) come from another source.
We also note that AGB models can provide reasonable fits to the abundances of these stars \cite{bisterzo12}. For HE0336+0113 and CS29528-028, the AGB models predict [Pb/Fe] $\sim 1.8$ and $\sim 3.5$, respectively. Our massive stars models predict [Pb/Fe] $<1$. Future determinations of the Pb abundance could help choosing between AGB and massive stars.

\ack

A.C. acknowledges funding from the Swiss National Science Foundation under grant P2GEP2\_184492. RH acknowledges support from the EU COST Action CA16117 (ChETEC) and from the WorldPremier International Research Centre Initiative (WPI Initiative), MEXT, Japan. 

\nocite{iliadis10a}
\nocite{taggart19}
\nocite{caughlan88}
\nocite{angulo99}
\nocite{cohen06}
\nocite{cohen13}
\nocite{prantzos90}
\nocite{best13}

\section*{References}

\bibliographystyle{jphysicsB}
\bibliography{biblio.bib}

\end{document}